\documentclass[prb,twocolumn,superscriptaddress,showpacs,amsmath,amssymb]{revtex4}
\usepackage{amsfonts}
\usepackage{bbm}
\usepackage{graphicx}
\usepackage{dcolumn}
\usepackage{bm}

\begin{document}
\title{Switch effect and 0-$\pi$ transition in Ising superconductor Josephson junctions}

\author{Qiang Cheng}
\affiliation{School of Science, Qingdao University of Technology, Qingdao, Shandong 266520, China}
\affiliation{International Center for Quantum Materials, School of Physics, Peking University, Beijing 100871, China}

\author{Qing-Feng Sun}
\email[]{sunqf@pku.edu.cn}
\affiliation{International Center for Quantum Materials, School of Physics, Peking University, Beijing 100871, China}
\affiliation{Collaborative Innovation Center of Quantum Matter, Beijing 100871, China}
\affiliation{CAS Center for Excellence in Topological Quantum Computation, University of Chinese Academy of Sciences, Beijing 100190, China}

\begin{abstract}
We theoretically study the Josephson current
in Ising superconductor$-$half-metal$-$Ising superconductor junctions.
By solving the Bogoliubov-de Gennes equations,
the Josephson currents contributed by the discrete Andreev levels
and the continuous spectrum are obtained.
For very short junctions, because
the direct tunneling of the Cooper pair dominates the Josephson current,
the current-phase difference relation is independent of the magnetization direction,
which is the same as the conventional superconductor-ferromagnet-superconductor junctions.
On the other hand, when the length of the half-metal is similar to or greater
than the superconducting coherence length, the spin-triplet Josephson effect occurs and dominates
the Josephson current. In this case, the current-phase difference relations show
the strong magnetoanisotropic behaviors with the period $\pi$.
When the magnetization direction points to the $\pm z$ directions,
the current is zero regardless of the phase difference.
However, the current has a large value when the magnetization direction is parallel to the junction plane, which leads to a perfect switch effect of the Josephson current.
Furthermore, we find that the long junctions can host both the $0$-state and $\pi$-state,
and the $0$-$\pi$ transitions can be achieved with the change of the magnetization direction.
The physical origins of the switch effect and $0$-$\pi$ transitions
are interpreted from the perspectives of the spin-triplet Andreev reflection,
the Ising pairing order parameter and the Ginzburg-Landau type of free energy.
In addition, the influences of the chemical potential, the magnetization magnitude and
the strength of the Ising spin-orbit coupling on the switch effect and $0$-$\pi$ transitions
are also investigated.
Furthermore, the two-dimensional Josephson junctions are also investigated and
we show that the spin-triplet Josephson effect can exist always.
These results provide a convenient way to control the Josephson critical current
and to adjust the junctions between the $0$-state and $\pi$-state by only rotating one magnetization.
\end{abstract}
\maketitle

\section{\label{sec1}Introduction}

Monolayer transition-metal dichalcogenides have been subjected to continuously
growing interest due to their potential applications in valleytronics\cite{Xiao,Cao}
and optoelectronics\cite{Wang,addSanchez}.
New physics is expected in the monolayer materials with the inversion symmetry breaking
and the strong Ising spin-orbit coupling (ISOC)\cite{Taguchi,Scharf}.
Recently, the superconductivity with the Ising pairing in atomically thin crystals
such as MoS$_{2}$ and NbSe$_{2}$ has been reported successively\cite{Lu,Saito,Xi,AddWang1,AddWang2}.
The in-plane upper critical field of the Ising superconductor (ISC) far exceeds
the Pauli paramagnetic limit because of the presence of ISOC\cite{Ilic}.
The superconducting phase diagrams and the topological properties\cite{Yuan,Sharma,Hsu,addAliabad}
of ISC are also theoretically studied in monolayer transition-metal dichalcogenides.
It is predicted that the topologically non-trivial phase can support
the chiral Majorana edge states\cite{Yuan}.

Researches on the Ising superconductivity open a new route for the superconducting spintronics.
For the conventional ferromagnet-superconductor junctions,
the conductance does not depend on the direction of magnetization.
When the ferromagnet becomes a half-metal (HM), the subgap conductance will vanish
since the Andreev reflection process is fully suppressed\cite{Andreev,Hirai,addzhu1}.
However, this is not the case of the HM-ISC junctions\cite{Zhou,Lv}.
When the direction of magnetization in HM is parallel to the plane of the HM-ISC junctions,
the equal-spin Cooper pair can be formed
and the spin-triplet Andreev reflection can occur\cite{Lv}, which will lead to the finite subgap conductance.
Recently, the magnetoanisotropic spin-triplet Andreev reflection in the ferromagnet-ISC junctions
is systematically studied by Lv $\emph{et al}$\cite{Lv}
using the nonequilibrium Green's function method.
A strong magnetoanisotropy with $\pi$-period is found,
which is different from the conventional magnetoanisotropic system with $2\pi$-period\cite{Julliere,Moodera,Butler,Zhuang}.
Even so, the study on the ISC Josephson junctions is still blank.

Magnetic Josephson junctions are another class of spintronic setup for investigating
the interplay between ferromagnetism and superconductivity\cite{Linder,Buzdin}.
It possesses practical applications in classical and quantum circuits.
The junctions can host the so-called $\pi$-state with the negative critical
current\cite{Bulaevskii,addcheng,addzhu2},
which is believed to be helpful in designing the noise-immune superconducting qubits\cite{Feofanov}.
The tunable $0$-$\pi$ junction is the essential component for information storage in the
superconducting computer\cite{Gingrich}.
The formation of the $\pi$-state in conventional superconductor-ferromagnet-superconductor
junctions is determined by the specific thickness of the interlayer\cite{Ryazanov,Kontos,addSperstad}.
Accordingly, the control of the $0$-$\pi$ transition can only be realized
through changing the size of the ferromagnet.
Another alternative structure is the junctions with the ferromagnetic multilayer and the $0$-$\pi$
transition is tuned by changing the relative orientation of magnetizations\cite{addHalterman,Eschrig,Houzet}.
However, the manipulations of the thickness and the relative orientation are
all inconvenient in the circuits. Achieving the easily controllable $0$-$\pi$ transition
in the simple Josephson structures remains an urgent problem to be solved in condensed matter physics.

In this paper, we study the Josephson current in the ISC-HM-ISC junctions
which are concise sandwich structures.
By solving the Bogoliubov-de Gennes (BdG) equations\cite{Gennes,BTK} for ISCs and HM
and applying suitable boundary conditions,
the Andreev levels and the Josephson current are obtained
for both the double-band and the single-band junctions.
When the length of HM, denoted by $L$, is far less than the superconducting coherence length $\xi_{0}$,
the direct tunneling of the Cooper pair dominates the Josephson current.
The current-phase difference relation is weakly dependent on the direction of the magnetization in the HM region.
On the other hand, when the length $L$ is similar to or greater than $\xi_{0}$,
the spin-triplet Josephson effect dominates the current.
Then the Josephson current exhibits a strong magnetoanisotropy with a period $\pi$.
The current is zero when the magnetization direction of HM points to the $\pm z$ directions.
However, it has a large value when the magnetization direction is parallel to the junction plane,
which leads to a perfect switch effect of the Josephson current.

Furthermore, the long ISC-HM-ISC junctions can host both the $0$-state and $\pi$-state,
and the $0$-$\pi$ transitions can be achieved with the change of the magnetization direction.
That is to say, the switch effect and the $0$-$\pi$ transitions
can be conveniently realized by rotating one magnetization
in ISC-HM-ISC junctions with a definite length $L$ of HM.
From the detailed dependencies, the $0$-$\pi$ transitions can be classified into two kinds
which are the slow one and the sudden one.
In addition, the effects of the chemical potential, the magnitude of magnetization and the strength of ISOC
on the spin-triplet Josephson current and the $0$-$\pi$ transitions are also investigated.
The physical origins of the spin-triplet
current and the $0$-$\pi$ transitions are clarified by introducing
the spin-triplet Andreev reflection mechanism,
transforming the superconducting order parameters and
constructing the Ginzburg-Landau type of free energy.

The organization of this paper is as follows.
We will start in Sec.~\ref{sec2} by demonstrating the Hamiltonian of the
ISC-HM-ISC junctions and deriving the expressions of the discrete and continuous
Josephson currents by using the BdG equations.
In Sec.~\ref{sec3}, we present the numerical results and discuss the spin-triplet Josephson current,
the $0$-$\pi$ transitions and the switch effect.
Sec.~\ref{sec4} provides the physical interpretations on the physical origin of our main results.
Sec.~\ref{sec5} discusses the two-dimensional properties of the Ising superconductor junctions.
Sec.~\ref{sec6} concludes this paper.
Some tedious derivation processes for the continuous Josephson current are relegated to Appendix.

\section{\label{sec2}Model and formalism}

\begin{figure}[!htb]
\includegraphics[width=1.0\columnwidth]{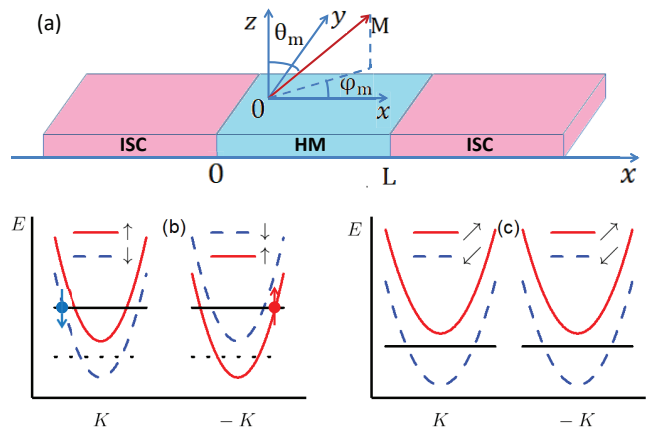}
\caption{(a) Schematic illustration of the ISC-HM-ISC junction.
The junction is in the $xy$-plane. The interface is located at $x=0$ and $x=L$.
The direction of magnetization $\bm{M}$ is depicted by the polar angle $\theta_{m}$
and the azimuthal angle $\varphi_{m}$.
(b) The energy bands near $\bm{K}$ and $-\bm{K}$ valleys for the normal phase of ISC.
The black solid lines and dashed lines indicate the Fermi energy
for the double-band case and the single-band case, respectively.
The red and blue arrows represent
two electrons with the opposite spin and opposite wave vector from different valleys,
which combine to form a Cooper pair.
(c) The energy bands for ferromagnet.
Here the Fermi energy (black solid lines) is across the energy band
with spin antiparallel to $\bm{M}$ only.}
\label{fig1}
\end{figure}

We consider the ISC-HM-ISC Josephson junctions as shown in Fig.1(a),
which are formed in a transition-metal dichalcogenide monolayer.
The left and right ISCs are semi-infinite while the length of the center HM is assumed as $L$.
The magnetization $\bm{M}$ in HM is specified by the polar angle $\theta_{m}$
and the azimuthal angle $\varphi_{m}$, i.e., ${\bm{M}}=M(\sin{\theta_{m}}\cos{\varphi_{m}},\sin{\theta_{m}}\sin{\varphi_{m}},\cos{\theta_{m}})$.
Its direction can be tuned continuously by a weak external field.

Due to the presence of two kinds of valleys ($\bm{K}$ and $\bm{-K}$) in the Brillouin zone,
the single-particle Hamiltonians for the normal phase of ISC are\cite{Zhou}
\begin{equation}
\hat{H}_{\pm}(k)=\frac{\hbar^2k^2}{2m}-\mu+\epsilon\beta\hat{\sigma}_{z}.\label{Ha1}
\end{equation}
Here, ${k}$ is the wave vector of electrons relative to the valleys $\pm\bm{K}$,
$\mu$ is the chemical potential,
$\epsilon=\pm$ is the valley index for $\pm\bm{K}$, $\beta$ is the strength of ISOC,
and $\hat{\sigma}_{z}$ is the Pauli matrix in the spin space.
In this section, we consider the one-dimensional Josephson junctions,
in which the wave vector ${k}$ only has one component.
The two-dimensional Josephson junctions will be studied in Sec.~\ref{sec5}.
The energy bands of the ISC's normal phase are schematically shown in Fig.1(b).
Here the spin sub-bands are split due to the ISOC.
At the $\bm{K}$ valley, the spin-up band has higher energy than the spin-down one,
but it is the opposite for the $-\bm{K}$ valley\cite{addKormanyos,addMattheiss}.
However, the ISC's normal phase still obeys the time-reversal symmetry
and the spin-rotation symmetry about the $z$ axis.
In Eq.(\ref{Ha1}), we have neglected the inter-valley scattering induced by impurity.
Since the valleys $\bm{K}$ and $\bm{-K}$ are located
at the corners of the Brillouin zone and are well separated,
the inter-valley scattering is very weak.

The BdG Hamiltonians for the superconducting region ($x<0$ or $x>L$) can be written as\cite{Zhou}
\begin{equation}
\check{H}_{BdG\pm}^{S}(k)=\left(
\begin{array}{cc}
\hat{H}_{\pm}(k)&\hat{\Delta}(k)\\
-\hat{\Delta}^{*}(-k)&-\hat{H}^{*}_{\mp}(-k)
\end{array}
\right),\label{HBdGS}
\end{equation}
in which $\hat{\Delta}({k})=\Delta e^{i\phi_{1(2)}} i\sigma_{y}$ is the superconducting order parameter
for the left (right) ISC with $\Delta$ the superconducting gap magnitude.
The phase difference $\phi$ of the left and right ISCs is defined as $\phi=\phi_{1}-\phi_{2}$.
For clarity, we will use $\mu_{s}$ and $\beta_{s}$ to denote the chemical potential and the strength of ISOC in ISC.
The Cooper pairs are formed by electrons with the opposite spin and opposite wave vector
from different valleys, as shown in Fig.1(b).
For $\mu_{s}>\beta_{s}$, ISC is a double-band superconductor and
for $\mu_{s}<\beta_{s}$, it is a single-band one [see Fig.1(b)].

The BdG Hamiltonians for the ferromagnetic region ($0<x<L$) are
\begin{equation}
\check{H}_{BdG\pm}^{F}(k)=\left(
\begin{array}{cc}
\hat{H}_{\pm}(k)+{\hat{\bm \sigma}}\cdot\bm{M}&0\\
0&-\hat{H}_{\mp}^{*}(-k)-\hat{\bm\sigma}^{*}\cdot\bm{M}
\end{array}
\right).\label{HBdGFM}
\end{equation}
We use $\mu_{f}$ and $\beta_{f}$ to denote the chemical potential and the strength of ISOC in this region.
In our model, $\beta_{f}$ is assumed to be negligible and will be set to zero.
Fig.1(c) schematically shows the energy bands of the ferromagnetic region.
Here the spin sub-bands are split due to the magnetization $\bm{M}$.
The band with spin parallel to $\bm{M}$ has higher energy than the antiparallel band
at both $\bm{K}$ and $-\bm{K}$ valleys.
In the ferromagnetic region, the time-reversal symmetry is broken.
If $M>\mu_{f}$, the Fermi energy is only across one sub-band as shown in Fig.1(c)
and this region becomes HM.

The total Josephson current can be divided into two parts,
the discrete current contributed by the discrete Andreev levels when the energy $|E|<\Delta$,
and the continuum current contributed by the continuous spectrum when $|E|>\Delta$.
Below we first derive the discrete current by solving the Andreev levels.
The wave functions of quasiparticles in each region can be obtained through
solving the BdG equations,
$\check{H}(-i\partial/\partial x)_{BdG\pm}\psi_{\pm}=E_{\pm}\psi_{\pm}$
with the substitution of $-i \partial/\partial x$ for ${k}$ in $\check{H}_{BdG\pm}({k})$.
The solution $\psi_{+}$ for ISCs is
\begin{eqnarray}
\psi_{+}(x<0) &=
 &c_{11}\xi_{e1}e^{-ik_{1}x}+c_{12}\xi_{e2}e^{-ik_{2}x} \nonumber\\
 && +d_{11}\xi_{h1}e^{ik_{1}x}+d_{12}\xi_{h2}e^{ik_{2}x},\label{ps0}
\end{eqnarray}
and
\begin{eqnarray}
\psi_{+}(x>L) & = &
 g_{11}\eta_{e1}e^{ik_{1}x}+g_{12}\eta_{e2}e^{ik_{2}x} \nonumber\\
 && +h_{11}\eta_{h1}e^{-ik_{1}x}+h_{12}\eta_{h2}e^{-ik_{2}x},\label{psL}
\end{eqnarray}
with the four-component vectors $\xi_{e1}=(ue^{i\phi_{1}/2},0,0,ve^{-i\phi_{1}/2})^{T}$, $\xi_{e2}=(0,ue^{i\phi_{1}/2},-ve^{-i\phi_{1}/2},0)^{T}$,
$\xi_{h1}=(ve^{i\phi_{1}/2},0,0,ue^{-i\phi_{1}/2})^{T}$ and $\xi_{h2}=(0,-ve^{i\phi_{1}/2},ue^{-i\phi_{1}/2},0)^{T}$.
One can obtain the vectors $\eta_{e1(2)}$ and $\eta_{h1(2)}$ by
substituting $\phi_{2}$ for $\phi_{1}$ in $\xi_{e1(2)}$ and $\xi_{h1(2)}$, respectively.
The coherent factors $u$ and $v$ are $u=\sqrt{(E+\Omega)/2E}$ and $v=\sqrt{(E-\Omega)/2E}$
with $\Omega=\sqrt{E^2-\Delta^2}$.
The wave vectors are expressed as $k_{1(2)}=\sqrt{2m(\mu_{s}-(+)\beta_{s})/\hbar^2}$
under the Andreev approximation\cite{Andreev}.
The solution $\psi_{-}$ can be found by interchanging
the two wave vectors $k_{1}$ and $k_{2}$ in $\psi_{+}$.
In $\psi_{-}$, we will use $c_{21(22)}$ and $d_{21(22)}$ to denote the coefficients
in front of $\xi_{e1(2)}$ and $\xi_{h1(2)}$ and will use $g_{21(22)}$ and $h_{21(22)}$
in front of $\eta_{e1(2)}$ and $\eta_{h1(2)}$, respectively.

The solution $\psi_{+}$ for the HM region ($0<x<L$) is
\begin{eqnarray}
 \psi_{+}(x) &= &
 f_{11}\chi_{e1}e^{iq_{e1}x}+f_{12}\chi_{e1}e^{-iq_{e1}x}+f_{13}\chi_{e2}e^{iq_{e2}x}\nonumber\\
&& +f_{14}\chi_{e2}e^{-iq_{e2}x} +f_{15}\chi_{h1}e^{iq_{h1}x}+f_{16}\chi_{h1}e^{-iq_{h1}x} \nonumber\\
 && +f_{17}\chi_{h2}e^{iq_{h2}x}+f_{18}\chi_{h2}e^{-iq_{h2}x},\label{pf}
\end{eqnarray}
where the four-component vectors are given by $\chi_{e1}=(\alpha_{1},\alpha_{2},0,0)^{T}$, $\chi_{e2}=(-\alpha_{2}^{*},\alpha_{1},0,0)^{T}$, $\chi_{h1}=(0,0,\alpha_{1},\alpha_{2}^{*})^{T}$
and $\chi_{h2}=(0,0,-\alpha_{2},\alpha_{1})^{T}$ with $\alpha_{1}=\cos(\theta_{m}/2)$ and $\alpha_{2}=\sin(\theta_{m}/2)e^{i\phi_{m}}$.
The wave vectors are expressed as
$q_{e(h)1}=\sqrt{2m(\mu_{f}-M)/\hbar^2}+(-)E/[2\sqrt{\hbar^2(\mu_{f}-M)/2m}]$ and $q_{e(h)2}=\sqrt{2m(\mu_{f}+M)/\hbar^2}+(-)E/[2\sqrt{\hbar^2(\mu_{f}+M)/2m}]$.
The solution $\psi_{-}$ possesses the same form of $\psi_{+}$ except that the coefficients
$f_{11}$, $f_{12}$, ... $f_{18}$ are replaced by $f_{21}$, $f_{22}$, ... $f_{28}$.

The boundary conditions at the ISC-HM interfaces are
\begin{align}
\psi_{\pm}(x=0^{-})=\psi_{\pm}(x=0^{+}),\label{eqbc1}\\
\psi^{'}_{\pm}(x=0^{-})=\psi^{'}_{\pm}(x=0^{+}),\label{eqbc2}\\
\psi_{\pm}(x=L^{-})=\psi_{\pm}(x=L^{+}),\label{eqbc3}\\
\psi^{'}_{\pm}(x=L^{-})=\psi^{'}_{\pm}(x=L^{+}).\label{eqbc4}
\end{align}
Eliminating the coefficients
$c_{11}$, $c_{12}$, $d_{11}$, $d_{12}$, $g_{11}$, $g_{12}$, $h_{11}$ and $h_{12}$
($c_{21}$, $c_{22}$, $d_{21}$, $d_{22}$, $g_{21}$, $g_{22}$, $h_{21}$ and $h_{22}$),
one will get the homogeneous linear equations
of $f_{11}$, $f_{12}$, ... and $f_{18}$ ($f_{21}$, $f_{22}$, ... and $f_{28}$).
Their coefficients construct a $8\times8$ matrix defined as $\Lambda_{1}$ ($\Lambda_{2}$).
The Andreev levels $E^{\pm}$ in the HM region are determined by\cite{Zagoskin}
\begin{equation}
\text{Det}[\Lambda_{1}(E^{+})]=0,
\label{eqd1}
\end{equation}
and
\begin{equation}
\text{Det}[\Lambda_{2}(E^{-})]=0.
\label{eqd2}
\end{equation}
The symbol $\text{Det}[\cdot\cdot\cdot]$ represents the determinant of a matrix.

The Josephson current contributed by the discrete Andreev levels is written as\cite{Bagwell,Zhang}
\begin{equation}
I_{d}=\frac{e}{\hbar}\sum_{\substack{n}}\left[\frac{dE_{n}^{+}}{d\phi}f(E_{n}^{+})+\frac{dE_{n}^{-}}{d\phi}f(E_{n}^{-})
 \right].
\label{eqid}
\end{equation}
Here, $f(E_{n}^{\pm})$ are the Fermi distribution functions.
The energies $E_{n}^{+}$ and $E_{n}^{-}$ denote two sets of discrete Andreev levels solved
from Eqs. (\ref{eqd1}) and (\ref{eqd2}), respectively.
The sum ensures the contributions from all Andreev levels are included.

Second, the Josephson current contributed by the continuous spectrum can be written as
\begin{equation}
I_{c}=\frac{e}{2h}\left(\int_{-\infty}^{-\Delta}+\int_{\Delta}^{\infty}\right)\left[\sum_{\lambda=\pm}(J_{e1}
^{\lambda}+J_{e2}^{\lambda}+J_{h1}^{\lambda}+J_{h2}^{\lambda})\right],
\end{equation}
with
\begin{align}
J_{e1(2)}^{\lambda}=\sum_{l=1,2}[(C_{e1(2)l}^{\lambda}-D_{e1(2)l}^{\lambda})
-(\tilde{C}_{e1(2)l}^{\lambda}-\tilde{D}_{e1(2)l}^{\lambda})],\label{Je12}\\
J_{h1(2)}^{\lambda}=\sum_{l=1,2}[(C_{h1(2)l}^{\lambda}-D_{h1(2)l}^{\lambda})
-(\tilde{C}_{h1(2)l}^{\lambda}-\tilde{D}_{h1(2)l}^{\lambda})]\label{Jh12},
\end{align}
where $C_{e1(2)l}^{\lambda}$ and $D_{e1(2)l}^{\lambda}$ describe the probabilities of transitions
as electron-like and hole-like quasiparticles respectively in the right ISC
when an electron-like quasiparticle characterized by $\xi_{e1(2)}$ is injected from the left ISC,
and $\tilde{C}_{e1(2)l}^{\lambda}$ and $\tilde{D}_{e1(2)l}^{\lambda}$ describe the probabilities of transitions
in the left ISC when the electron-like quasiparticle is injected from the right ISC.
$C_{h1(2)l}^{\lambda}$, $D_{h1(2)l}^{\lambda}$, $\tilde{C}_{h1(2)l}^{\lambda}$
and $\tilde{D}_{h1(2)l}^{\lambda}$ describe the similar processes
when a hole-like quasiparticle is injected.
The definition and derivation of these probabilities can be found in the Appendix.

The total Josephson current is expressed as
\begin{equation}
I=I_{d}+I_{c},
\end{equation}
which is a function of the phase difference $\phi$, the chemical potentials $\mu_{s}$ and $\mu_{f}$,
the ISOC strength $\beta_{s}$, the magnitude and direction of the magnetization $\bm{M}$ and the length $L$ of the HM region.

In this paper, we focus our attentions on the ISC-HM-ISC Josephson junctions with $\mu_{f}<M$
(i.e. the central region is HM with the complete spin polarization).
The temperature is taken as zero.
Since $f({E_{n}^{\pm}})$ will become step functions at the zero temperature,
the Andreev levels with $E_{n}^{\pm}>0$ do not contribute to the Josephson current $I$.
In the following calculations, we also take a specific energy $\mu_{0}=100\Delta$
as the unit of other energies such as $\mu_{s}$, $\beta_{s}$, $\mu_{f}$ and $M$.
The wave vector defined by $\mu_{0}$ is $k_{F}=\sqrt{2m\mu_{0}/\hbar^2}$.
The reciprocal of $k_{F}$ is the unit of the length $L$.
The superconducting coherence length is defined as $\xi_{0}=\hbar v_{Fs}/\pi\Delta$
with $v_{Fs}$ the Fermi velocity in ISCs.
Since ISCs obey the spin-rotation symmetry about the $z$ axis,
the Josephson current of the ISC-HM-ISC junctions will not depend on the azimuthal angle $\varphi_{m}$.

\section{\label{sec3}Results and discussions}
\subsection{\label{sec3-1} Double-band junctions}

\begin{figure}[!htb]
\includegraphics[width=1.0\columnwidth]{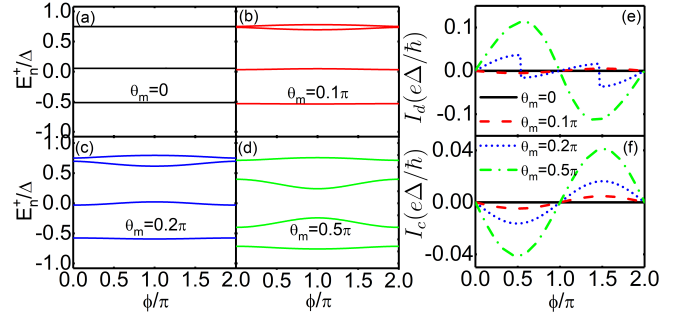}
\caption{
(a)-(d) The discrete Andreev levels $E_{n}^{+}$ for $\theta_{m}=0,0.1\pi,0.2\pi$ and $0.5\pi$, respectively.
(e) The discrete Josephson current $I_{d}$ and (f) the continuum Josephson current $I_{c}$
versus the phase difference $\phi$ for the different $\theta_{m}$.
The related parameters are $k_{F}L=100$, $\mu_{s}=1.3$, $\beta_{s}=1.1$, $\mu_{f}=1.0$ and $M=1.2$.}
\label{fig2}
\end{figure}

First of all, we study the double-band junctions with $\mu_{s}>\beta_{s}$.
Fig.2(a-d) shows the Andreev levels $E_{n}^{+}$
with the different polar angle $\theta_{m}$ of magnetization.
The HM length is $k_{F}L=100$ which is about the coherence length $\xi_{0}$ of ISCs.
Here, we do not show the Andreev levels $E_{n}^{-}$ for simplicity
since the equality $E_{n}^{-}=-E_{n}^{+}$ always holds.
For $\theta_{m}=0$, all Andreev levels $E_{n}^{+}$ are flat and they are independent of the
superconducting phase difference $\phi$ [see Fig.2(a)].
In fact, the magnetization $\bm{M}$ in this situation is in the $+z$ direction
and there only exist electrons with their spin pointing to the $-z$ direction in the HM region.
However, it needs spin-up (the $+z$ direction) and spin-down (the $-z$ direction) electrons to form Cooper pairs.
Therefore, there is a lack of the effective coupling between the states in HM and
Cooper pairs in ISCs. At present, ISCs only play the parts of the confinement potentials which cause the flat Andreev levels.
As the polar angle $\theta_{m}$ rises from $0$,
the Andreev levels gradually move down and start to depend on the phase difference $\phi$ [see Fig.2(b-d)]
due to the appearance of the spin-up electrons in HM.
When $\theta_{m}$ rises to $0.5\pi$, the Andreev levels $E_{n}^{+}$
distribute symmetrically about $E=0$.
In particular, the Andreev levels are significantly dependent on $\phi$
at $\theta_{m}=0.5\pi$ [see Fig.2(d)].

From these discrete Andreev levels in Fig.2(a-d) and by using Eq.(\ref{eqid}), the discrete Josephson current $I_d$ can be obtained as shown in Fig.2(e).
We also show the continuum Josephson current $I_c$ in Fig.2(f).
As the polar angle $\theta_{m}=0$, both the discrete current and the continuum current
are zero regardless of the phase difference $\phi$.
In this case, there only exist the spin-down electrons in the HM region [see Fig.1(c)].
The absence of spin-up electrons will forbid the occurrence of the Andreev reflection\cite{addsun1},
which results in the Josephson current being zero ($I_d=I_c=0$).
When $\theta_{m}$ deviates from zero, the nonzero currents, including the discrete one and the continuum one, begin to appear, which are the spin-triplet Josephson currents associated with the spin-triplet Andreev reflection.
The physical description of the spin-triplet Josephson currents is given in Sec. \ref{sec4}A.

Now, we discuss the discrete and continuum Josephson currents in detail.
Both $I_d$ and $I_c$ are strongly magnetoanisotropic and the current-phase difference
relations depend on the polar angle $\theta_m$. This is different from the
conventional superconductor-ferromagnet-superconductor junctions where the
current-phase difference relations are independent of $\theta_m$.
As the polar angle $\theta_{m}$ rises from $0$ to $0.5\pi$,
the amplitude of the continuum current $I_{c}$ is increased and the curves keep the sinusoidal form [see Fig.2(f)].
Here $I_c$ is negative when $0<\phi<\pi$.
However, the discrete current $I_{d}$ experiences a complicated evolution as shown in Fig.2(e).
The critical discrete current
for $\theta_{m}=0.1\pi$ is negative while that for $\theta_{m}=0.5\pi$ is positive.
With the increase of $\theta_{m}$, the amplitude of $I_{d}$ also increases.
The amplitude reaches its biggest value at $\theta_{m}=0.5\pi$.
In addition, for $\theta_{m}=0.2\pi$, there are two jumps of current near $\phi=0.5\pi$ and $1.5\pi$.
These behaviors of $I_{d}$ can be understood from the Andreev levels $E_{n}^{+}$ in Fig.2(a-d).
For $\theta_{m}=0.1\pi$, there is only one Andreev level below the Fermi energy $E_F=0$ [see Fig.2(b)],
which level is concave and leads to the negative critical value.
For $\theta_{m}=0.2\pi$, the second lowest Andreev level crosses with $E_F=0$ [see Fig.2(c)],
which induces the jumps of $I_{d}$.
Furthermore, for $\theta_{m}=0.5\pi$, the second lowest level is below $E_F$ and it is convex,
which will provide the main contribution to $I_{d}$ and bring about the positive critical current.

\begin{figure}[!htb]
\includegraphics[width=1.0\columnwidth]{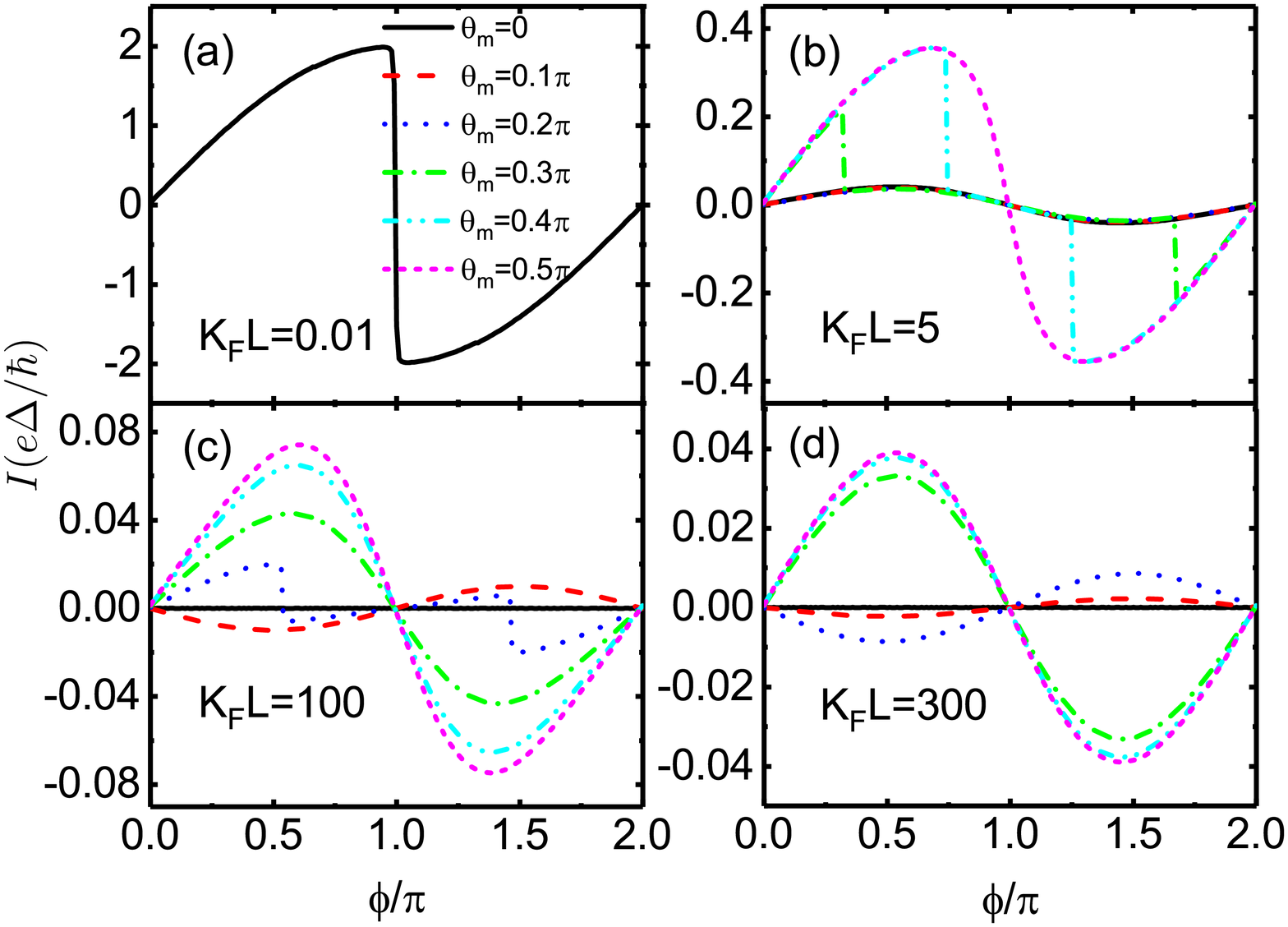}
\caption{The total current $I$ as a function of the phase difference $\phi$
with $\theta_{m}=0,0.1\pi,0.2\pi,0.3\pi,0.4\pi$ and $0.5\pi$ for (a) $k_{F}L=0.01$,
(b) $k_{F}L=5$, (c) $k_{F}L=100$ and (d) $k_{F}L=300$.
Other parameters have the same values as those in Fig.2.}
\label{fig2}
\end{figure}

Next we focus on the total Josephson current $I$.
Fig.3 shows the total current $I$ as a function of the phase difference $\phi$
for different values of the HM's length $L$.
Firstly, in usual, the discrete current $I_{d}$ is much larger than
the continuum current $I_c$ [Fig.2(e) and 2(f)].
Consequently, the discrete current $I_{d}$ dominates the shapes of the total current [see Fig.2(a) and 3(c)].
Secondly, the total current strongly relies on the length $L$ of HM.
For $L\sim0$ as shown in Fig.3(a), the current remains unchanged when the magnetization is rotated.
This is because ISCs are directly coupled with each other.
The current-phase difference relation reduces to that of ISC-ISC junctions,
and $I$ can almost reach the biggest value $2e\Delta/\hbar$.
In this case, the Josephson current $I$ originates from the direct tunneling of the Cooper pair.
As the length $L$ increases, the current $I$ gradually decreases.
When $k_F L=5$ ($L\ll\xi_{0}$ is still satisfied), the direct tunneling of the Cooper pair becomes weak
but the current is still finite even for $\theta_{m}=0$ [see Fig.3(b)].
Meanwhile, the $\theta_{m}$-dependence of the current starts to emerge,
which means the occurrence of the spin-triplet Josephson effect.
When $L\sim\xi_{0}$ as shown in Fig.3(c), the strongly magnetoanisotropic Josephson current is exhibited.
The current $I$ for $\theta_{m}=0$ is zero regardless of the phase difference $\phi$,
because the direct tunneling of the Cooper pair disappears.
But the spin-triplet Josephson effect by the multiple Andreev reflection occurs, which leads to
a large current at $\theta_m =0.5\pi$.
The current possesses the ``on-off" property when one rotates the magnetization
from $\theta_{m}\ne0$ to zero.
This switch effect is an important result of the ISC-HM-ISC junctions.
Another important effect of our junctions is the $0$-$\pi$ transition.
The negative critical current for $\theta_{m}=0.1\pi$ indicates the formation of the $\pi$-state
with the current-phase difference relation $\sim\sin(\phi+\pi)$. Different from the $0$-state,
the minimum of the free energy is now achieved at $\phi=\pi$ not $\phi=0$.\cite{Bulaevskii}
The two important effects manifest themselves more clearly when $L>\xi_{0}$ as shown in Fig.3(d).

\begin{figure}[!htb]
\includegraphics[width=1.0\columnwidth]{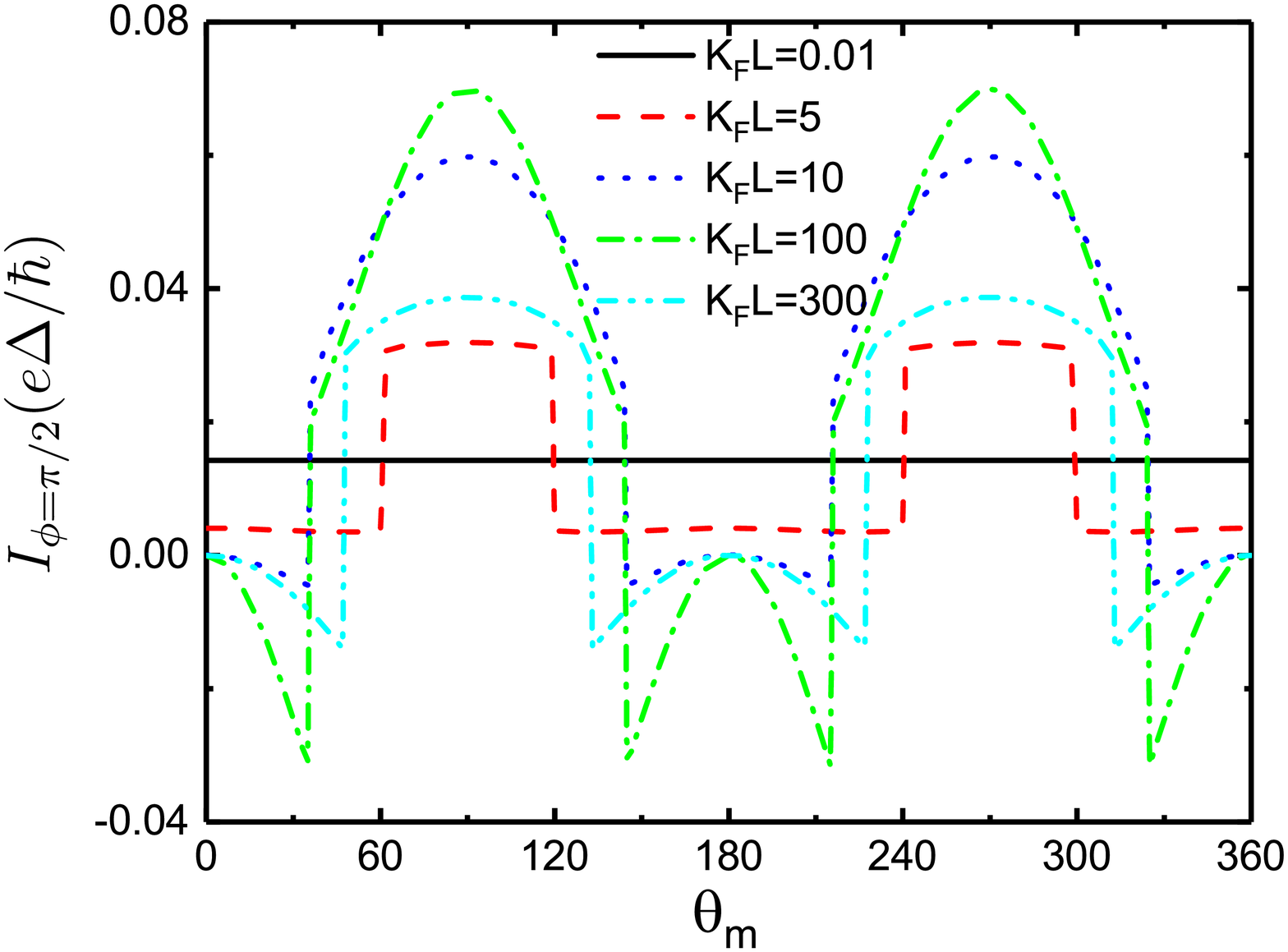}
\caption{
The total current $I$ with $\phi=0.5\pi$ as a function of the polar angle $\theta_{m}$
for $k_{F}L=0.01,5,10,100$ and $300$.
The currents for $k_{F}L=0.01$ (the black solid line) and $k_{F}L=5$ (the red dashed line)
have been taken as $1/100$ and $1/10$ of their real values.
Other parameters have the same values as those in Fig.2.}
\end{figure}

The detailed $\theta_{m}$-dependence of the total current $I$ at $\phi=0.5\pi$ can be found in Fig.4.
The current displays periodic variations with a period of $\pi$.
This is distinct from the conventional superconductor-ferromagnet-superconductor junctions,
where the Josephson current is independent of the direction of the magnetization $\bm{M}$.
Within one period, the current is symmetric about $\theta_{m}=90^{\circ}$ or $\theta_{m}=270^{\circ}$
which indicates $I(\theta_{m})=I(\pi-\theta_{m})$.
In order to explain this symmetry,
we introduce the rotation operation around the $x$ axis with the rotating angle $180^{\circ}$.
The operation is defined as the unitary matrix $\mathcal{M}_{x}=\text{diag}(m_{x},m_{x}^{*})$
with $m_{x}=i\sigma_{x}$.
Under this transformation, the Hamiltonians $\check{H}_{BdG\pm}^{F}$ with $\theta_{m}$
are changed to $\check{H}_{BdG\mp}^{F}$ with $\pi-\theta_{m}$.
In other words, the direction of $\bm{M}$ in HM is rotated from $\theta_{m}$ to $\pi-\theta_{m}$.
Simultaneously, the Hamiltonians $\check{H}_{BdG\pm}^{S}$ are changed to $\check{H}_{BdG\mp}^{S}$.
If we denote the current associated with $\check{H}_{BdG\pm}$ by $I_{\pm}$, then $I_{\pm}(\theta_{m})=I_{\mp}(\pi-\theta_{m})$ is satisfied. The total current $I$, as the sum of $I_{+}$ and $I_{-}$,
meets the invariance $I(\theta_{m})=I(\pi-\theta_{m})$.
In addition, since the spin-triplet effect depends only on the magnetization component in the $xy$ plane not the component along the $z$ direction,
we also obtain $I(\theta_{m}) = I(\pi-\theta_{m})$.
Moreover, considering that ISCs have the spin-rotation symmetry about the $z$ axis
and the spherical coordinates $(\theta_{m}+\pi,\varphi_{m})$ and $(\pi-\theta_{m},\varphi_{m}+\pi)$
are equative, we have $I(\theta_{m}+\pi)=I(\pi-\theta_{m})$.
By combining $I(\theta_{m}) = I(\pi-\theta_{m})$ and $I(\theta_{m}+\pi)=I(\pi-\theta_{m})$,
it brings about the $\pi$-periodicity Josephson current straightforwardly.

From Fig.4, the following conclusions can also be drawn.
The current $I$ for $k_{F}L=0.01$ does not rely on the polar angle $\theta_{m}$,
because the direct tunneling of the Cooper pair dominates the current.
When $L\ll\xi_{0}$ (e.g. $k_{F}L=5$),
the current is always greater than zero and approximates the shape of a square wave,
which implies that the ISC-HM-ISC junctions locate the $0$-state regardless of $\theta_{m}$.
With the increase of $L$, the current strongly depends on the polar angle $\theta_m$
because of the emergence of the magnetoanisotropic spin-triplet Josephson effect.
Furthermore, the $\pi$-state can be formed
even for a short junction (see the curve for $k_{F}L=10$).
Now, the ISC-HM-ISC junctions can host the $0$-state or $\pi$-state
by tuning the direction of the magnetization $\bm{M}$.
For the greater values of $k_{F}L$, the $\pi$-state is either more pronounced ($k_{F}L=100$)
or formed in a wider angle range of $\theta_{m} $($k_{F}L=300$).
When the current reaches its negative maximum, a sudden transition from the $\pi$ state to the $0$ state will happen.
Actually, the sudden transition is always accompanied with the formation of the $\pi$-state.
The physical explanation of the sudden transition between the $0$ state and the $\pi$ state will be given in Sec. {\ref{sec4}}B.

\begin{figure}[!htb]
\includegraphics[width=1.0\columnwidth]{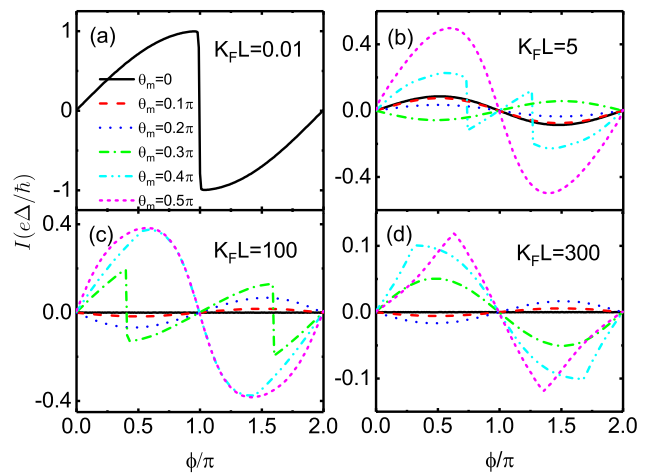}
\caption{
The total current $I$ as a function of the phase difference $\phi$
with $\theta_{m}=0,0.1\pi,0.2\pi,0.3\pi,0.4\pi$ and $0.5\pi$
for (a) $k_{F}L=0.01$, (b) $k_{F}L=5$, (c) $k_{F}L=100$ and (d) $k_{F}L=300$.
The related parameters are $\mu_{s}=1.0$, $\beta_{s}=1.1$, $\mu_{f}=1.0$ and $M=1.2$.}
\end{figure}

\subsection{\label{sec3-2} Single-band junctions}
Now, we turn to the single-band case with $\mu_{s}<\beta_{s}$.
Plotted in Fig.5 shows the current-phase difference relations for $\mu_{s}=1.0$ and $\beta_{s}=1.1$.
For $L\sim0$ in Fig.5(a), the current is irrespective of the polar angle $\theta_{m}$
due to the direct tunneling of the Cooper pair. It can almost reach the biggest value $e\Delta/\hbar$, half of the value for the double-band junctions [see Fig.3(a)].
For $L\ll\xi_{0}$ in Fig.5(b), the current at $\theta_{m}=0$ is not equal to zero and
it also depends on $\theta_{m}$. In this case, the direct tunneling of the Cooper pair and the
spin-triplet Josephson current coexist.
These results are analogous to those for the double-band case.
However, the $0$-$\pi$ transition in the single-band junctions can occur for shorter length $L$
than that of the double-band case.
For the single-band junctions with $k_{F}L=5$, the $0$-$\pi$ transition has appeared [see Fig.5(b)].
Actually, there are two types of $0$-$\pi$ transitions as $\theta_{m}$
is increased from $0$ to $0.5\pi$.
One takes place slowly near $\theta_{m}=45^{\circ}$ and the other occurs suddenly
near $\theta_{m}=67^{\circ}$, which have been shown clearly in Fig.6.
For $L\sim\xi_{0}$ in Fig.5(c), the current is zero at $\theta_{m}=0$.
Now, the spin-triplet Josephson current dominates the total current.
The switch effect and the $0$-$\pi$ transition can occur when one raises $\theta_{m}$ from zero.
For $L>\xi_{0}$ in Fig.5(d), the switch effect and the $0$-$\pi$ transition still exist
and new current-phase difference relations like a triangular wave can be obtained.

\begin{figure}[!htb]
\includegraphics[width=1.0\columnwidth]{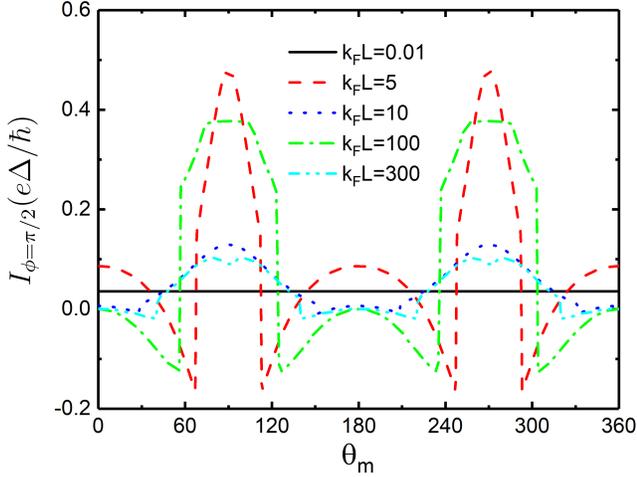}
\caption{
The total current $I$ with $\phi=0.5\pi$ as a function of the polar angle $\theta_{m}$
for $k_{F}L=0.01,5,10,100$ and $300$.
The current for $k_{F}L=0.01$ (the black solid line) has been taken as $1/20$ of its real value.
Other parameters have the same values as those in Fig.5.}
\end{figure}

Fig.6 shows the $\theta_{m}$-dependence of the total current at $\phi=0.5\pi$ for the single-band junctions.
The current exhibits the $\pi$-periodicity $I(\theta_{m})=I(\pi+\theta_{m})$
and the relation $I(\theta_{m}) =I(\pi-\theta_{m})$,
which are the same as those for the double-band case.
The current for the HM's length $L\sim0$ is a nonzero constant
due to the direct tunneling of the Cooper pair.
For $L\ll\xi_{0}$ with $k_{F}L=5$, the spin-triplet Josephson current begins to appear,
which leads to the result that the current is magnetoanisotropic (i.e. the current depends on $\theta_m$),
but the current $I$ at $\theta_{m}=0$ is still a nonzero positive value
by the tunneling of the Cooper pair.
As $\theta_{m}$ increases from $0$ to $0.5\pi$, the positive $I$ gradually decreases and changes into
a negative value, then $I$ suddenly jumps to a large positive value.
As a result, there are two types of $0$-$\pi$ transitions, the slow one and the sudden one.
For larger values of $L$, the direct tunneling of the Cooper pair is very weak.
Thus, $I$ is zero at $\theta_{m}=0$ and the slow $0$-$\pi$ transition disappears.
However, the current $I$ is large at $\theta_{m}=0.5\pi$ due to the spin-triplet Josephson effect.
By tuning the direction of the magnetization,
the Josephson critical current can easily be regulated,
and the switch effect is activated.
Moreover, for the single-band junctions, the increased length of HM
is not always beneficial to the formation of the $\pi$-state.

Next, we will take $k_{F}L=5$ as an example to
discuss the two types of $0$-$\pi$ transitions from the angle of Andreev levels.
We first consider the sudden $0$-$\pi$ transition.
The discrete Andreev levels $E_{n}^{+}$ and $E_{n}^{-}$
as functions of the polar angle $\theta_{m}$ for $k_{F}L=5$ and $\phi=0.5\pi$ are drawn in Fig.7(a).
There are four intersections between the levels and $E_F=0$.
The positions of the intersections give the values of $\theta_{m}$
for the sudden transitions in Fig.6.
In order to clear up how the transitions happen,
we take the first intersection point and mark it by $A$.
On the left of the point A, the level $E_{2}^{-}<0$ and
contributes to the Josephson current according to Eq.(\ref{eqid}),
while on the right of the point A, $E_{2}^{-}>0$ and the level $E_{1}^{+}<0$ contributes to the current.
The derivatives of $E_{2}^{-}$ and $E_{1}^{+}$ with respect to $\phi$ are negative and positive,
respectively [see Fig.7(b)], so the current suddenly changes its sign
when $\theta_{m}$ passes the point A,
which brings about the occurrence of a sudden $0$-$\pi$ transition.
Furthermore, in Sec. {\ref{sec4}}B, we give
the physical explanation of the sudden $0$-$\pi$ transition from the spin-triplet Cooper pairs.

\begin{figure}[!htb]
\includegraphics[width=1.0\columnwidth]{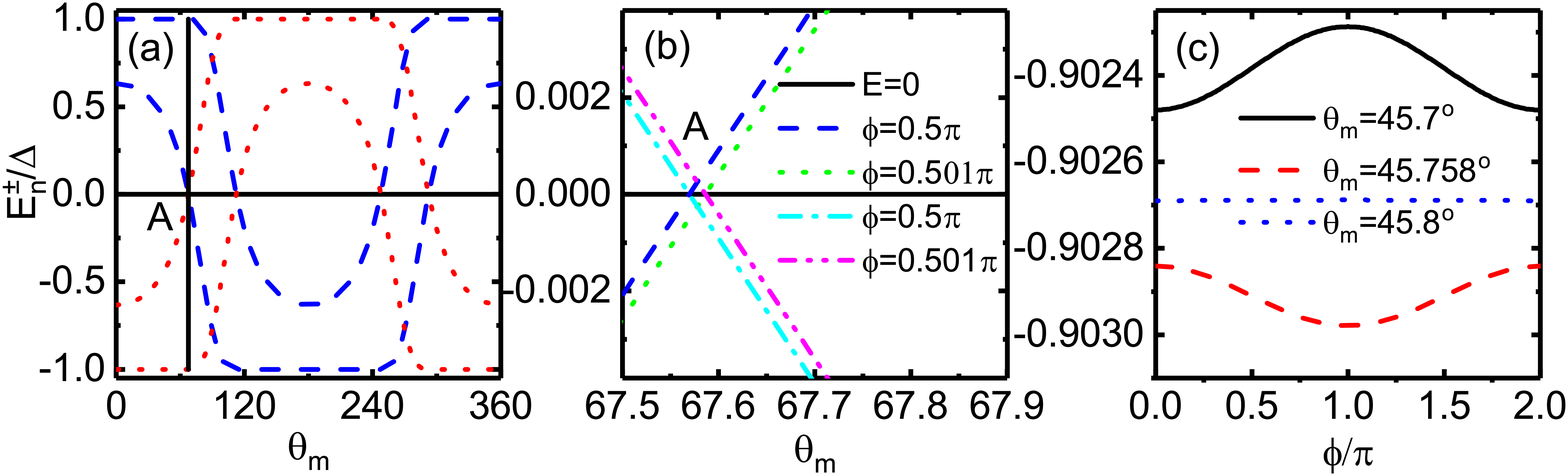}
\caption{
(a) The Andreev levels $E_{n}^{+}$ (the blue dashed curves)
and $E_{n}^{-}$ (the red dotted curves) as functions of $\theta_{m}$ for $k_{F}L=5$
and $\phi=0.5\pi$. The black solid line represents $E_F=0$.
The symbol $A$ denotes the intersection point between the Andreev levels and $E_F=0$.
(b) The enlarged figure in the vicinity of the point $A$ in (a)
with $\phi=0.5\pi$ and $0.501\pi$.
(c) The Andreev levels versus the phase difference $\phi$ for $\theta_{m}=45.7^{\circ}<\theta_m^c$,
$45.758^{\circ}=\theta_m^c$ and $45.8^{\circ}>\theta_m^c$.
Other parameters have the same values as those in Fig.6.}
\end{figure}

Then we consider the slow $0$-$\pi$ transition.
From Fig.6, the slow transition arises at $\theta_{m}= \theta_{m}^c$
($\theta_{m}^c \approx 45.758^{\circ}$).
Fig.7(c) shows the Andreev level-phase difference relations for $\theta_{m}<\theta_{m}^c$,
$\theta_{m}=\theta_{m}^c$ and $\theta_{m}>\theta_{m}^c$.
These three curves respectively are concave, flat and convex, and
their slopes at $\phi=0.5\pi$ are positive, zero and negative.
As a result, it gives rise to a slow evolution of the junctions
from the $0$-state to the $\pi$-state as $\theta_{m}$ increases
from less than $\theta_{m}^c$ to greater than $\theta_{m}^c$.
Due to the periodicity and the relation $I(\theta_m)=I(\pi-\theta_m)$,
the other three points for the slow $0$-$\pi$ transition in Fig.6 can also be obtained.

\begin{figure}[!htb]
\includegraphics[width=1.0\columnwidth]{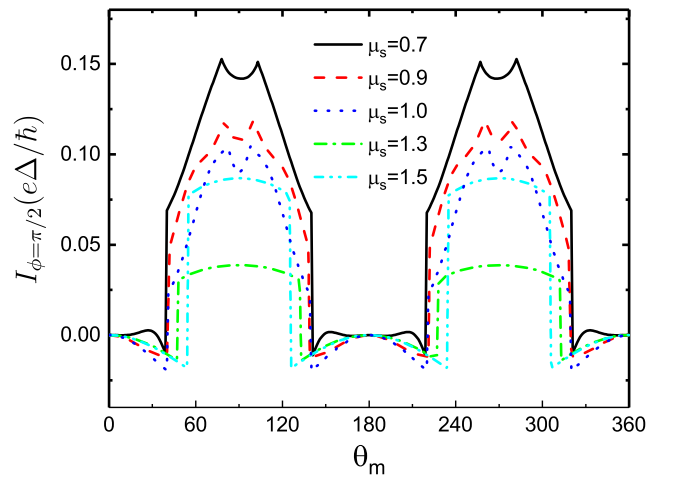}
\caption{The total current $I$ with $\phi=0.5\pi$ as a function of $\theta_{m}$
for various values of $\mu_{s}$. The parameters are $k_{F}L=300$, $\beta_{s}=1.1$, $\mu_{f}=1.0$ and $M=1.2$.}
\end{figure}

\begin{figure}[!htb]
\includegraphics[width=1.0\columnwidth]{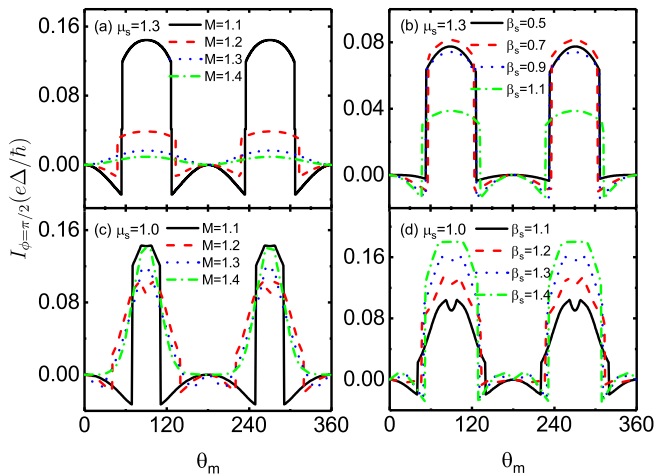}
\caption{
The total current $I$ with $\phi=0.5\pi$ as a function of $\theta_{m}$
for (a) $\mu_{s}=1.3$, $\beta_{s}=1.1$ and various values of $M$,
(b) $\mu_{s}=1.3$, $M=1.2$ and various values of $\beta_{s}$,
(c) $\mu_{s}=1.0$, $\beta_{s}=1.1$ and various values of $M$,
and (d) $\mu_{s}=1.0$, $M=1.2$ and various values of $\beta_{s}$.
Other parameters are $k_{F}L=300$ and $\mu_{f}=1.0$.}
\end{figure}

\subsection{\label{sec3-3}
Effects of system parameters on the spin-triplet Josephson current}

Let us investigate the effect of the chemical potential $\mu_{s}$ on the spin-triplet Josephson current.
Fig.8 shows the total current $I$ versus the polar angle $\theta_m$ for the different $\mu_{s}$.
Here the HM's length $L$ is taken as $k_{F}L=300$, where the direct tunneling of the Cooper pair disappears.
The current $I$ exhibits a strong magnetoanisotropy
for both the double-band junctions ($\mu_s>\beta_s)$ and single-band junctions ($\mu_s<\beta_s$)
due to the spin-triplet Josephson effect.
The current is always zero at $\theta_m=0$ and it has the large value at $\theta_m=0.5\pi$.
As a result, the switch effect can be achieved for all $\mu_s$.
Furthermore, both the $0$-state and $\pi$-state can appear,
and the transition between them is always sudden regardless of the $\mu_s$.
With the increase of $\mu_{s}$, the angle range realizing the $\pi$-state becomes larger.
Because of the presence of the sudden $0$-$\pi$ transition,
one can conveniently adjust between the $0$-state and $\pi$-state by tuning the polar angle $\theta_m$.

Finally, we study the effects of the magnetization magnitude $M$ and ISOC strength $\beta_{s}$
on the Josephson current (see Fig.9).
The current $I$ exhibits a strong magnetoanisotropy for all $M$ and $\beta_{s}$,
where $I=0$ at $\theta_m=0$ and $I$ is large at $\theta_m=0.5\pi$
because of the spin-triplet Josephson effect. Thus the switch effect always holds.
On the other hand, the sudden $0$-$\pi$ transition is gradually weakened
as the magnetization magnitude $M$ increases. When $M$ is much larger than $\mu_f$,
the $\pi$-state disappears and there is no $0$-$\pi$ transition
both for the double-band junctions [see Fig.9(a)] and single-band junctions [see Fig.9(c)].
In contrast, the $\pi$-state can survive regardless of the ISOC strength $\beta_s$,
and the sudden $0$-$\pi$ transition can be present both for the double-band junctions [Fig.9(b)] and single-band junctions [Fig.9(d)]. In addition, there exist current dips around $\theta_{m}=90^{\circ}$ and $\theta_{m}=270^{\circ}$
for the single-band junctions [see Fig.8 and Fig.9(c-d)].
The dips express the deviation of the current-phase difference relation
from the sinusoidal form which has been seen in Fig.5(d).
For $\theta_{m}\sim0.5\pi$ and $L>\xi_{0}$,
the current no longer obtains its maximum value at $\phi=0.5\pi$ but at $\phi >0.5\pi$.
The dip magnitude is almost independent of $\mu_{s}$ as plotted in Fig.8.
When $M$ is raised or $\beta_{s}$ is reduced,
the dips will gradually fade away as given in Fig.9(c) and (d).
However, for the double-band junctions, there is no current dip
and the current is always the largest at $\theta_m=0.5\pi$ [see Fig.8 and Fig.9].
It is consistent with the current-phase difference relations presented in Fig.3(d).

\section{\label{sec4}Physical interpretations}
\subsection{\label{sec4-1} Switch effect}

Now, we explain the origin of the switch effect. In other words, we clarify how the spin-triplet Josephson current comes into being when $\theta_{m}\ne0$.
When $\theta_{m}$ deviates from zero, the magnetization in HM is no longer collinear to
the spin-quantization axis (the $+z$ direction) of ISCs [see Fig.1(c)].
The spin wave function of electrons in HM can be written as
the superposition of spin-up and spin-down relative to the $z$ axis.
As a result, the spin-triplet Andreev reflection becomes possible\cite{Lv}.
Taking $\theta_{m}=0.5\pi$ as an example,
the spin of the electrons in HM all points to the $-x$ direction.
The $-x$ spin state can split up into the spin-up (the $+z$ direction) and spin-down (the $-z$ direction) states.
Considering that a spin-up electron in the HM region moves forward and reaches the
right HM-ISC interface, the spin-triplet Andreev reflection occurs,
where the spin-up electron is reflected back as
a spin-down hole in HM and a Cooper pair is injected into the right ISC.
Then, when the spin-down hole reaches the left ISC-HM interface,
the Andreev reflection occurs again with a spin-up electron reflected back and
a Cooper pair annihilated in the left ISC.
The above process repeats again and again, and the Josephson current flows through the
ISC-HM-ISC junction.

In addition, the aforementioned process can also be regarded as that a Cooper pair is injected
from the left ISC, splits into two electrons with their spin pointing to the $-x$ direction
in the central HM region, and combines into the Cooper pair in the right ISC again,
which brings the Josephson current.
Note that the spin of the two electrons in HM is in the $-x$ direction,
i.e. they are in a spin-triplet state with the total spin $S=1$ and $S_x=-1$.
Hence, this is a spin-triplet Josephson effect.
Since the Cooper pair in the ISCs has the spin-triplet component,
the spin-triplet Andreev reflection can occur in the HM-ISC interface and the spin-triplet
Josephson current can flow through the ISC-HM-ISC junctions.
This is essentially different from the conventional superconductor-HM-superconductor junctions
where the Andreev reflection can not occur and the Josephson current disappears.

\subsection{\label{sec4-2} $0$-$\pi$ transitions}

Next, we explain the origin of the $0$-$\pi$ transitions.
Due to the presence of the ISOC, the Pauli matrices $\hat{\sigma}_x$
and $\hat{\sigma}_y$ are not commutative with the ISC's Hamiltonians in
Eqs.(\ref{Ha1}) and (\ref{HBdGS}).
Thus, the total spin ${\bm{S}}$ is not a good quantum number
and the wave function of Cooper pairs in ISC has both the
spin-singlet and spin-triplet components.
Following Ref.[\onlinecite{Zhou}],
the spin-triplet pairing correlation can be obtained, which is
\begin{eqnarray}
\Delta d_z({\bm{k}},E) \hat{\sigma}_z i \hat{\sigma}_y =\Delta d_z({\bm{k}},E)
\left(\begin{array}{cc}
0&1\\
1&0
\end{array}\right),\label{STP1}
\end{eqnarray}
where $d_z({\bm{k}},E) = 2\epsilon \beta \xi_k /[(\Delta^2+\xi_k^2-E^2)^2
+2\beta^2(\Delta^2 -\xi_k^2-E^2) +\beta^4]$ with $\xi_k =\frac{\hbar^2 k^2}{2m} -\mu$.
Here $\epsilon =\pm$ is the valley index for $\pm\bm{K}$.
The parameters $\Delta$, $\beta$, $\mu$ and ${\bm{k}}$ are the same as those in
the Hamiltonians (\ref{Ha1}) and (\ref{HBdGS}).
In Eq.(\ref{STP1}), the spin-quantization axis is at the $z$ direction.
If we chose the direction of the magnetization in HM as the quantization axis,
the spin-triplet paring correlation changes to the following form,
\begin{equation}
d_{z}(\textbf{k})
\left(\begin{array}{cc}
-\sin{\theta_{m}}&\cos{\theta_{m}}\\
\cos{\theta_{m}}&\sin{\theta_{m}}
\end{array}\right).\label{ISOP}
\end{equation}
This order parameter possesses the same structure
as that for the spin-triplet superconductor without ISOC.

The wave function in the spin-triplet superconductor is described
by the $\bm{d}$-vector.\cite{Balian}
We consider the spin-triplet superconductor$-$ferromagnet$-$spin-triplet superconductor junctions
with $\bm{d}\parallel\hat{z}$, i.e., $\bm{d}=\tilde{d}_{z}(\bm{k})\hat{z}$ with the orbital part $\tilde{d}_{z}(\bm{k})$. The order parameter in the superconductors is
\begin{equation}
\left(\begin{array}{cc}
0&\tilde{d}_{z}(\bm{k})\\
\tilde{d}_{z}(\bm{k})&0
\end{array}\right).
\end{equation}
The form of the order parameter also depends on the choice of the spin-quantization axis.
If we chose the direction of the magnetization in ferromagnet
as the quantization axis as we have done for ISC,
the order parameter will bear the same form as that in Eq.(\ref{ISOP}) for ISC except for the different factors $d_{z}(\bm{k})$ and $\tilde{d}_{z}(\bm{k})$.

The crucial term in the current-phase difference relations,
which is responsible for the formation of $0$-$\pi$ transitions
in the spin-triplet Josephson junctions, is as follows\cite{Brydon,Bujnowski},
\begin{eqnarray}
I\propto -\cos{2\theta_{m}}\sin{\phi}.\label{CPR}
\end{eqnarray}
For $0\le\theta_{m}<\pi/4$, $I\propto-\sin{\phi}$ corresponds to the $\pi$ state,
while for $\pi/4<\theta_{m}\le\pi/2$, $I\propto\sin{\phi}$ corresponds to the $0$ state.
The sign change of $I$ at $\theta_{m}=\pi/4$ leads to the $0$-$\pi$ transition.
Taking the influences of other structure parameters into account,
the transition angle will acquire a deviation from $\pi/4$ \cite{Bujnowski}.
Note, although the $\bm{d}$-vectors in Refs.[\onlinecite{Brydon,Bujnowski}]
is taken along the $x$ axis,
the above analyses with $\bm{d}\parallel\hat{z}$ are also consistent.

The $0$-$\pi$ transition at $\pi/4$ also applies to the ISC-HM-ISC Josephson junctions
due to the wave function of Cooper pairs in ISC
having the spin-triplet components.
Because the influences of the chemical potentials,
the ISOC strength, the magnetization magnitude and the length of HM,
the $0$-$\pi$ transition angle deviates from $\pi/4$ as shown in Figs.4, 6, 8 and 9,
but it is always around $\pi/4$.

In addition, the current-phase difference relation in Eq.(\ref{CPR})
can also be derived through constructing the Ginzburg-Landau type of free energy
for the ISC-HM-ISC Josephson junctions.
Generally, for the magnetic Josephson junctions with the spin-triplet paring characterized
by $\bm{d}$-vectors, the free energy can always be constructed with the magnetization $\bm{M}$
and the $\bm{d}$-vectors.
The selection rules for the lowest order current in the spin-triplet Josephson junctions
have been well explained using the constructed free energy \cite{Cheng1,Cheng2}.
The constructed terms can not only demonstrate the characteristics of the current-phase difference relation
but also directly express the interplay of ferromagnetism and superconductivity.

Now, we turn to the ISC-HM-ISC Josephson junctions.
Assuming the spin-quantization axis along the magnetization $\bm{M}$,
the $\bm{d}$-vector for the left (right) ISC is
\begin{eqnarray}
\bm{d}_{l(r)}=d_{z}(\bm{k})(\sin{\theta_{m}},0,\cos{\theta_{m}})e^{\phi_{1(2)}},
\end{eqnarray}
according to the order parameter in Eq.$(\ref{ISOP})$.
We postulate that the following two terms will contribute to the free energy,
\begin{eqnarray}
[(\bm{d}_{l}\cdot\bm{M})(\bm{M}\cdot\bm{d}^{*}_{r})+H.c.],\label{T1}
\end{eqnarray}
and
\begin{eqnarray}
[(\bm{d}_{l}\times\bm{M})\cdot(\bm{M}\times\bm{d}_{r}^{*})+H.c.].\label{T2}
\end{eqnarray}
The symbol ``$*$" denotes the conjugation operation which guarantees the $U(1)$ gauge invariance of the free energy.
Substituting $\bm{d}_{l(r)}$ and $\bm{M}=(0,0,M)$ into Eqs.(\ref{T1}) and (\ref{T2}), we get the free energy $F\propto (\cos^{2}{\theta_{m}}-\sin^{2}{\theta_{m}})\cos{\phi}$. The Josephson current, as the
derivative of the free energy with respect to $\phi$, is proportional to $-\cos{2\theta_{m}}\sin{\phi}$ which is just the term in Eq.(\ref{CPR}). The term is consistent with the relation $I(\theta_{m})=I(\pi-\theta_{m})$ and the periodicity $I(\theta_{m})=I(\pi+\theta_{m})$.

\section{\label{sec5}Two-dimensional ISC-HM-ISC junctions}
\subsection{\label{sec5-1} Formalism}

In the previous sections, the one-dimensional ISC-HM-ISC junctions are studied only.
In this section, we discuss the properties of the tow-dimensional ISC-HM-ISC junctions.
In this situation, the size along the $y$ direction of the left ISC, center HM region, and
right ISC are finite [see Fig.1(a)].
Then the Hamiltonian $\hat{H}_{\pm}$ in Eq.(\ref{Ha1}) changes into:
\begin{equation}
\hat{H}_{\pm}({\bm k})=\frac{\hbar^2 {\bm k}^2}{2m}-\mu+\epsilon\beta\hat{\sigma}_{z}.\label{Ha2D}
\end{equation}
Compared with the one-component wave vector $k$ in Eq.(\ref{Ha1}),
here the wave vector has two components with ${\bm k}=(k_x,k_y)$.
The BdG Hamiltonians of the ISCs and HM regions for the two-dimensional ISC-HM-ISC junctions
are the same as Eqs.(\ref{HBdGS}) and (\ref{HBdGFM}),
and only the $\hat{H}_{\pm}(k)$ in them needs to be replaced by $\hat{H}_{\pm}({\bm k})$
in Eq.(\ref{Ha2D}).
We consider the periodic boundary condition at the $y$ direction
and the wave vector $k_y$ is a good quantum number which is conserved in the scattering process.

For a given $k_{y}$, the wave functions in ISCs and HM can be derived by solving the BdG equations $\check{H}(-i\partial/\partial x,k_{y})_{BdG\pm}\psi_{\pm}=E_{\pm}\psi_{\pm}$ with the substitution of $-i\partial/\partial x$ for $k_{x}$ in $\check{H}_{BdG\pm}(k_{x},k_{y})$. The obtained wave functions have the same form as those in Eqs.({\ref{ps0}})-(\ref{pf}). However, the wave vectors need to be rewritten as $k_{1(2)}=\sqrt{2m(\mu_{s}-(+)\beta_{s})/\hbar^2-k_{y}^2}$, $q_{e(h)1}=\sqrt{2m(\mu_{f}-M)/\hbar^2-k_{y}^2}+(-)E/[2\sqrt{\hbar^2(\mu_{f}-M-\hbar^2k_{y}^2/2m)/2m}]$ and $q_{e(h)2}=\sqrt{2m(\mu_{f}+M)/\hbar^2-k_{y}^2}+(-)E/[2\sqrt{\hbar^2(\mu_{f}+M-\hbar^2k_{y}^2/2m)/2m}]$. The $k_{y}$-dependent coefficients in the wave functions and the $k_{y}$-dependent Andreev levels $E_{\pm}$ in HM can be determined by the conditions Eqs.(\ref{eqbc1})-(\ref{eqbc4}) and Eqs. ($\ref{eqd1}$) and ($\ref{eqd2}$), respectively.

For the given $k_{y}$, the contribution to the Josephson current along the $x$ axis
is expressed as $I(k_{y})=[I_{d}(k_{y})+I_{c}(k_{y})]\cos{\theta}$
with the incident angle $\theta=\sin^{-1}(k_{y}/\sqrt{\mu_{s}+\beta_{s}})$
which is the angle between the wave vector $\bm k=(k_{2}, k_{y})$ and the $x$ axis. Next, we will use the dimensionless $k_{y}$ normalized by the wave vector $k_{F}$ defined in Sec.~\ref{sec2}.
There exists a critical wave vector $k_{yc}=\sqrt{\mu_{s}+\beta_{s}}$.
Only these wave vectors $k_{y}$ with $k_{y}<k_{yc}$ contribute to the Josephson current.
When $k_y>k_{yc}$, $I(k_{y})$ is zero.
If we assume the junction size along the $y$ direction is $W$,
the normalized wave vector $k_{y}$ can be written as $k_{y}=2\pi n/(k_{F}W)$ with $n$
an integer number under the periodic boundary condition\cite{Titov}.
The two-dimensional current $I$ will be the sum of $I(k_{y})$ over $k_{y}$.

\subsection{\label{sec5-2} Results and discussions}

\begin{figure}[!htb]
\includegraphics[width=1.0\columnwidth]{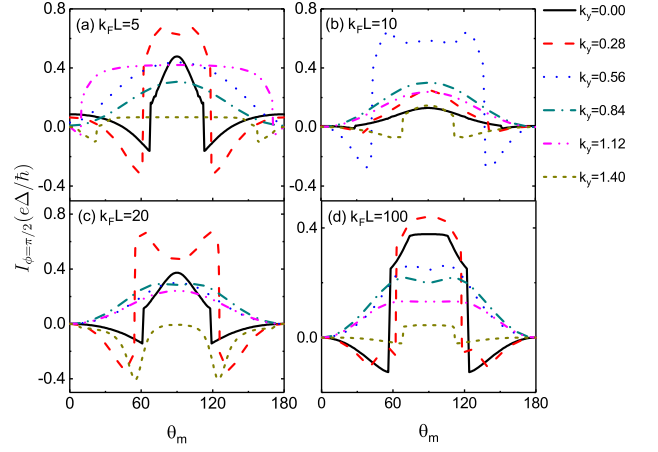}
\caption{The Josephson current $I(k_y)$ with different $k_{y}$ as a function of $\theta_{m}$
for (a) $k_{F}L=5$, (b) $k_{F}L=10$, (c) $k_{F}L=20$ and (d) $k_{F}L=100$.
Other parameters are $\mu_{s}=1.0$, $\beta_{s}=1.1$, $\mu_{f}=1.0$ and
 $M=1.2$.}\label{p1}
\end{figure}

Firstly, we consider the properties of narrow junctions with the small value of $W$.
If the width $W$ is smaller than a critical width $W_c$ ($k_{F}W_c= 2\pi/k_{yc}$),
only the wave vector $k_{y}=0$ contributes to the Josephson current.
This is just the one-dimensional case that we have discussed in the previous sections.
In this case, the spin-triplet Josephson effect occurs.
The Josephson current strongly depends on the magnetization angle $\theta_{m}$ in the HM
with the magnetoanisotropic period being $\pi$,
which leads to the perfect switch effect and $0$-$\pi$ transitions.
In the Ref.[\onlinecite{Zhou}], the junction parameters are taken
as $\mu_{s}=4.0\Delta$ and $\beta_{s}\approx 2.7\Delta$.
By using these parameters, the critical width $W_c$ is about $\xi_{0}$
with $\xi_{0}$ being the superconducting coherence length.

\begin{figure}[!htb]
\includegraphics[width=1.0\columnwidth]{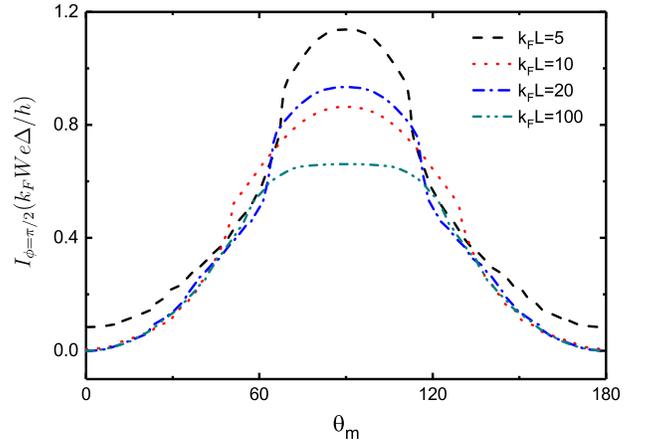}
\caption{
The Josephson current for the two-dimensional junctions along the $x$ axis
as a function of $\theta_{m}$ for $k_{F}L=5, 10, 20$ and $100$.
Other parameters have the same values as those in Fig.10.}\label{p1}
\end{figure}

Secondly, we consider the properties of wider junctions with the width $W >W_c$
and the contribution of the wave vector $k_{y}\ne 0$ to the Josephson current.
Fig.10 show the $k_{y}$-dependences of the Josephson current $I(k_y)$ as
a function of the polar angle $\theta_{m}$ of the magnetization in HM.
Here the junction parameters are chosen as $\mu_{s}=1.0$ and $\beta_{s}=1.1$,
which are the same as those in Figs.5 and 6.
In Fig.10, both the short and the long junctions are considered.
It is the most obvious feature that the spin-triplet Josephson effect
still takes effect for all values of $k_{y}$, leading to that
the Josephson current strongly depends on the polar angle $\theta_{m}$.
In other words, the Josephson current still exhibits a strong magnetoanisotropy
and the magnetoanisotropic period is $\pi$.
The current is very small at $\theta_{m}=0$ and generally acquires a large value at $\theta_{m}=0.5\pi$.
For the short junctions with $k_{F}L=5$, the Josephson current at $\theta_{m}=0$
has a small non-zero value due to the direct tunneling of Cooper pairs [see Fig.10(a)].
For the longer junctions, the current is vanishing at $\theta_{m}=0$ [see Fig.10(b-d)].
But at $\theta_{m}=0.5\pi$ the current generally has a large value regardless of
the length $L$ and wave vector $k_y$.
So the switch effect persists for all junctions.
On the other hand, the $0$-$\pi$ transition can keep for some wave vectors $k_{y}\ne0$,
e.g. see the curves with $k_{y} = 0.28$ in Figs.10(a, c and d)
and the curves with $k_{y}= 0.56$ and $1.40$ in Fig.10(b),
but for others, the $0$-$\pi$ transition is weak with the small negative current
or vanishing.

Thirdly, we consider the junctions with large enough $W$.
In this situation, the wave vector $k_{y}$ tends to be continuous.
The sum over $k_{y}$ will turn into the integral over $k_{y}$.
The Josephson current in the two-dimensional ISC-HM-ISC junctions
after integral of the normalized $k_{y}$ is given by $I=\frac{k_{F}W}{2\pi}\int{I(k_{y})}dk_{y}$.
Fig.11 shows the Josephson current versus the polar angle $\theta_{m}$
for the two-dimensional junctions at the superconducting phase difference $\phi =\pi/2$.
The spin-triplet Josephson effect still survives for both the short and the long junctions.
The magnetoanisotropy and its period are not affected by the dimensionality.
For $k_{F}L=5$, the small non-zero value of the Josephson current at $\theta_{m}=0$ originates
from the direct tunneling of Cooper pairs.
For the longer junctions, the spin-triplet effect dominates the Josephson current.
The Josephson current is zero at $\theta_{m}=0$
and has the maximum value at $\theta_{m}=0.5\pi$.
So the switch effect can well persist for the two-dimensional Josephson junctions.
On the other hand, the $0$-$\pi$ transition no longer exists under such circumstance.

\section{\label{sec6}Summary}
In conclusion, we systematically study the Josephson effect
in the sandwich structure consisting of Ising superconductors and half-metal.
By using the Bogoliubov-de Gennes equations,
the discrete Josephson current is calculated through solving the Andreev levels
and the continuous Josephson current is expressed as the composition of transition probabilities.
For different values of the length $L$ of half-metal, the total Josephson current shows different characteristics.
When the length is very short, the direct tunneling of the Cooper pair dominates
the Josephson current which is independent of the direction of the magnetization.
However, for the long junctions, the spin-triplet Josephson current dominates,
which exhibits a strong magnetoanisotropy with the period $\pi$.
The spin-triplet Josephson current completely disappears
as the magnetization direction points to the $\pm z$ directions,
but it has the large value as the magnetization direction is parallel to the junction plane.
Thus the junctions can work as a switch of the Josephson current.
Furthermore, with the change of the magnetization direction,
the junctions can host both the $0$-state and $\pi$-state.
At a special magnetization direction, a sudden $0$-$\pi$ transition occurs.
This provides a convenient experimental way to regulate the $0$-state and $\pi$-state
by tuning the magnetization direction.
In addition, the influences of the chemical potential, the strength of magnetization and
the Ising spin-orbit coupling are also investigated,
which help to specify suitable parameters for the experimental realization of the $\pi$-state
in a simple structure.
The mechanism for the spin-triplet Andreev reflection, the exotic order parameter in Ising superconductors and the Ginzburg-Landau type of free energy are explored,
which are responsible for the formations of the switch effect and the $0$-$\pi$ transitions.
At last, we show that the spin-triplet Josephson effect can well survive
in the two-dimensional junctions
and the Josephson current is strongly magnetoanisotropic with a period $\pi$
always.

\section*{ACKNOWLEDGMENTS}
This work was financially supported by National Key R and D Program of China (2017YFA0303301),
NBRP of China (2015CB921102), NSF-China under Grants Nos. 11574007 and 11447175,
the Strategic Priority Research Program of Chinese Academy of Sciences (XDB28000000)
and the Natural Science Foundation of Shandong Province under Grants No. ZR2017QA009.

\section*{APPENDIX}

\setcounter{equation}{0}
\renewcommand{\theequation}{A.\arabic{equation}}

Consider that an electron-like quasiparticle characterized by $\xi_{e1}$ is injected from the left ISC.
Following the BdG equation $\check{H}_{BdG+}(-i\nabla_{\bm{r}})\psi_{+}=E_{+}\psi_{+}$,
the wave function $\psi_{+}$ in the superconducting region is represented as
\begin{equation}
\begin{split}
\psi_{+}(x<0)=&\xi_{e1}e^{ik_{1}x}+a_{e11}^{+}\xi_{h1}e^{ik_{1}x}+a_{e12}^{+}\xi_{h2}e^{ik_{2}x}\\
+&b_{e11}^{+}\xi_{e1}e^{-ik_{1}x}+b_{e12}^{+}\xi_{e2}e^{-ik_{2}x},
\end{split}
\end{equation}
and
\begin{equation}
\begin{split}
\psi_{+}(x>L)=&c_{e11}^{+}\xi_{e1}e^{ik_{1}x}+c_{e12}^{+}\xi_{e2}e^{ik_{2}x}\\
+&d_{e11}^{+}\xi_{h1}e^{-ik_{1}x}+d_{e12}^{+}\xi_{h2}e^{-ik_{2}x}.
\end{split}
\end{equation}
The wave function in the ferromagnetic region is
\begin{equation}
\begin{split}
\psi_{+}(0<x<L)=&f_{11}^{+}\chi_{e1}e^{iq_{e1}x}+f_{12}^{+}\chi_{e1}e^{-iq_{e1}x}\\
+&f_{13}^{+}\chi_{e2}e^{iq_{e2}x}+f_{14}^{+}\chi_{e2}e^{-iq_{e2}x}\\
+&f_{15}^{+}\chi_{h1}e^{iq_{h1}x}+f_{16}^{+}\chi_{h1}e^{-iq_{h1}x}\\
+&f_{17}^{+}\chi_{h2}e^{iq_{h2}x}+f_{18}^{+}\chi_{h2}e^{-iq_{h2}x}.
\end{split}
\end{equation}
Here, $a_{e11}^{+}$ and $a_{e12}^{+}$ are the Andreev reflection coefficients,
$b_{e11}^{+}$ and $b_{e12}^{+}$ are the normal reflection coefficients,
$c_{e11}^{+}$ and $c_{e12}^{+}$ are the transition coefficients for electron-like quasiparticles
and $d_{e11}^{+}$ and $d_{e12}^{+}$ are the transition coefficients for hole-like quasiparticles.
The subscript $e$ in the coefficients denotes the injection of an electron-like quasiparticle.
The superscript $+$ denotes that the scattering process is described by the wave function
$\psi_{+}$ solved from the equation $\check{H}_{BdG+}(-i\nabla_{\bm{r}})\psi_{+}=E_{+}\psi_{+}$.

Applying the boundary conditions Eqs.(\ref{eqbc1})-(\ref{eqbc4}),
the analytic expressions of these coefficients can be derived.
The probabilities for the reflection and transition processes can be defined as
\begin{align}
A_{e11}^{+}=&\vert{a_{e11}^{+}}\vert^{2},~~~
A_{e12}^{+}=\text{Re}\left[\frac{k_{2}}{k_{1}}\right]\vert{a_{e12}^{+}}\vert^{2},\label{eqa}\\
B_{e11}^{+}=&\vert{b_{e11}^{+}}\vert^{2},~~~
B_{e12}^{+}=\text{Re}\left[\frac{k_{2}}{k_{1}}\right]\vert{b_{e12}^{+}}\vert^{2},\label{eqb}\\
C_{e11}^{+}=&\vert{c_{e11}^{+}}\vert^{2},~~~
C_{e12}^{+}=\text{Re}\left[\frac{k_{2}}{k_{1}}\right]\vert{c_{e12}^{+}}\vert^{2},\label{eqc}\\
D_{e11}^{+}=&\vert{d_{e11}^{+}}\vert^{2},~~~
D_{e12}^{+}=\text{Re}\left[\frac{k_{2}}{k_{1}}\right]\vert{d_{e12}^{+}}\vert^{2}.\label{eqd}
\end{align}
The defined quantities above satisfy the conservation of probability,
\begin{equation}
\sum_{l=1,2}(A_{e1l}^{+}+B_{e1l}^{+}+C_{e1l}^{+}+D_{e1l}^{+})=1.
\end{equation}

When an electron-like quasiparticle characterized by $\xi_{e1}$ is injected from the right ISC,
we can derive the coefficients and define the probabilities in a similar way. They are
\begin{align}
\tilde{A}_{e11}^{+}=&\vert{\tilde{a}_{e11}^{+}}\vert^{2},~~~
\tilde{A}_{e12}^{+}=\text{Re}\left[\frac{k_{2}}{k_{1}}\right]\vert{\tilde{a}_{e12}^{+}}\vert^{2},\label{eqta}\\
\tilde{B}_{e11}^{+}=&\vert{\tilde{b}_{e11}^{+}}\vert^{2},~~~
\tilde{B}_{e12}^{+}=\text{Re}\left[\frac{k_{2}}{k_{1}}\right]\vert{\tilde{b}_{e12}^{+}}\vert^{2},\label{eqtb}\\
\tilde{C}_{e11}^{+}=&\vert{\tilde{c}_{e11}^{+}}\vert^{2},~~~
\tilde{C}_{e12}^{+}=\text{Re}\left[\frac{k_{2}}{k_{1}}\right]\vert{\tilde{c}_{e12}^{+}}\vert^{2},\label{eqtc}\\
\tilde{D}_{e11}^{+}=&\vert{\tilde{d}_{e11}^{+}}\vert^{2},~~~
\tilde{D}_{e12}^{+}=\text{Re}\left[\frac{k_{2}}{k_{1}}\right]\vert{\tilde{d}_{e12}^{+}}\vert^{2}.\label{eqtd}
\end{align}
Actually, the quantities in Eqs.(\ref{eqta})-(\ref{eqtd}) can easily be found from Eqs.(\ref{eqa})-(\ref{eqd})
by the transformation $\phi\rightarrow-\phi$.

$C_{e11}^{+}$, $C_{e12}^{+}$, $D_{e11}^{+}$, $D_{e12}^{+}$, $\tilde{C}_{e11}^{+}$, $\tilde{C}_{e12}^{+}$, $\tilde{D}_{e11}^{+}$ and $\tilde{D}_{e12}^{+}$ in Eqs.(\ref{eqc}), (\ref{eqd}), (\ref{eqtc})
and $(\ref{eqtd})$ are just the quantities appearing in Eq.(\ref{Je12}) in the main text.
The other twenty four probability coefficients in $J_{e2}^{+}$, $J_{h1}^{+}$ and $J_{h2}^{+}$
can be solved by considering the following six processes described by $\psi_{+}$:
an electron-like (a hole-like) quasiparticle characterized by $\xi_{e2}$ ($\xi_{h1}$ or $\xi_{h2}$)
is injected from the left and the right ISC.
Applying the same method to the eight processes described by $\psi_{-}$,
the thirty two probability coefficients in $J_{e1}^{-}$, $J_{e2}^{-}$, $J_{h1}^{-}$ and $J_{h2}^{-}$
will be obtained in a similar way.

\end{document}